\def\blankline{\vskip \baselineskip}
\def\DF{{\tenrm DF}}

\def\half{{\leavevmode\kern.1em\raise.5ex\hbox{\the\scriptfont0 1}\kern-.1em
/\kern-.1em\lower.25ex\hbox{\the\scriptfont0 2}}} 
\def\quarter{{\leavevmode\kern.1em\raise.5ex\hbox{\the\scriptfont0 1}\kern-.1em
/\kern-.1em\lower.25ex\hbox{\the\scriptfont0 4}}}

\def\spose#1{\hbox to 0pt{#1\hss}} 
\def\ltsim{\mathrel{\spose{\lower 3pt\hbox{$\mathchar"218$}}
     \raise 2.0pt\hbox{$\mathchar"13C$}}}
\def\gtsim{\mathrel{\spose{\lower 3pt\hbox{$\mathchar"218$}}
     \raise 2.0pt\hbox{$\mathchar"13E$}}}
\def\gtlt{\mathrel{\spose{\lower 3pt\hbox{$\mathchar"13E$}}
     \raise 3pt\hbox{$\mathchar"13C$}}}

\def\pmb#1{\setbox0=\hbox{$#1$}%
  \kern-0.25em\copy0\kern-\wd0
  \kern.05em\copy0\kern-\wd0
  \kern-0.025em\raise.0433em\box0}

\def\today{\count99=\day
           \ifnum\count99>20 \count98=\day
                             \divide\count98 by 10
                             \multiply\count98 by 10
                             \advance\count99 by -\count98 \fi
           \number\day\ifcase\count99 th\or st\or nd\or rd\else th\fi
           ~\ifcase\month none\or January\or February\or March\or April\or
                  May\or June\or July\or August\or September\or October\or
                  November\or December\fi
           ~\number\year}

\long\def\Ignore#1{\relax}%

\newdimen\digitwidth
\setbox0=\hbox{0}
\digitwidth=\wd0
%
%

\def\eg{{\it e.g.}}
\def\etal{{\it et al.}}

\newcount\linespacingstep
\newcount\linespacing
\linespacingstep=1
\def\multiplelines{\linespacing=\linespacingstep
   \advance\linespacing by 2
   \multiply\normalbaselineskip by \linespacing
   \advance\normalbaselineskip by 2pt 
   \divide\normalbaselineskip by 3}

\font\fiverm=cmr5	\font\sixrm=cmr6	\font\sevenrm=cmr7
\font\eightrm=cmr8 	\font\ninerm=cmr9	\font\tenrm=cmr10
\font\twelverm=cmr12 	

\font\fivei=cmmi5 	\font\sixi=cmmi6 	\font\seveni=cmmi7
\font\eighti=cmmi8 	\font\ninei=cmmi9 	\font\teni=cmmi10
\font\twelvei=cmmi12

\font\fivesy=cmsy5  	\font\sixsy=cmsy6 	\font\sevensy=cmsy7
\font\eightsy=cmsy8 	\font\ninesy=cmsy9 	\font\tensy=cmsy10
\font\magnifiedtensy=cmsy10 at 12pt

\font\fivebf=cmbx5  	\font\sixbf=cmbx6 	\font\sevenbf=cmbx7
\font\eightbf=cmbx8 	\font\ninebf=cmbx9 	\font\tenbf=cmbx10
\font\twelvebf=cmbx12

	\font\eightit=cmti8 	\font\nineit=cmti9
\font\tenit=cmti10	\font\twelveit=cmti12

\font\eightsl=cmsl8 	\font\ninesl=cmsl9	\font\tensl=cmsl10
\font\twelvesl=cmsl12

\font\eighttt=cmtt8 	\font\ninett=cmtt9 	\font\tentt=cmtt10
\font\twelvett=cmtt12

\font\tenex=cmex10

\def\eightpoint{\def\rm{\fam0\eightrm}
\textfont0=\eightrm\scriptfont0=\sixrm\scriptscriptfont0=\fiverm
\textfont1=\eighti\scriptfont1=\sixi\scriptscriptfont1=\fivei
\textfont2=\eightsy\scriptfont2=\sixsy\scriptscriptfont2=\fivesy
\textfont3=\tenex\scriptfont3=\tenex\scriptscriptfont3=\tenex
\textfont\itfam=\eightit\def\it{\fam\itfam\eightit}
\textfont\slfam=\eightsl\def\sl{\fam\slfam\eightsl}
\textfont\ttfam=\eighttt\def\tt{\fam\ttfam\eighttt}
\textfont\bffam=\eightbf\scriptfont\bffam=\sixbf
\scriptscriptfont\bffam=\fivebf\def\bf{\fam\bffam\eightbf}
\normalbaselineskip=9pt
\ifnum\linespacingstep>1\multiplelines\fi
\setbox\strutbox=\hbox{\vrule height7pt depth2pt width0pt}
\let\sc=\sixrm\let\big=\eightbig\normalbaselines\rm}

\def\ninepoint{\def\rm{\fam0\ninerm}
\textfont0=\ninerm\scriptfont0=\sixrm\scriptscriptfont0=\fiverm
\textfont1=\ninei\scriptfont1=\sixi\scriptscriptfont1=\fivei
\textfont2=\ninesy\scriptfont2=\sixsy\scriptscriptfont2=\fivesy
\textfont3=\tenex\scriptfont3=\tenex\scriptscriptfont3=\tenex
\textfont\itfam=\nineit\def\it{\fam\itfam\nineit}
\textfont\slfam=\ninesl\def\sl{\fam\slfam\ninesl}
\textfont\ttfam=\ninett\def\tt{\fam\ttfam\ninett}
\textfont\bffam=\ninebf\scriptfont\bffam=\sixbf
\scriptscriptfont\bffam=\fivebf\def\bf{\fam\bffam\ninebf}
\normalbaselineskip=11pt
\ifnum\linespacingstep>1\multiplelines\fi
\setbox\strutbox=\hbox{\vrule height8pt depth3pt width0pt}
\let\sc=\sevenrm\let\big=\ninebig\normalbaselines\rm}

\def\tenpoint{\def\rm{\fam0\tenrm}
\textfont0=\tenrm\scriptfont0=\sevenrm\scriptscriptfont0=\fiverm%
\textfont1=\teni\scriptfont1=\seveni\scriptscriptfont1=\fivei%
\textfont2=\tensy\scriptfont2=\sevensy\scriptscriptfont2=\fivesy%
\textfont3=\tenex\scriptfont3=\tenex\scriptscriptfont3=\tenex%
\textfont\itfam=\tenit\def\it{\fam\itfam\tenit}%
\textfont\slfam=\tensl\def\sl{\fam\slfam\tensl}%
\textfont\ttfam=\tentt\def\tt{\fam\ttfam\tentt}%
\textfont\bffam=\tenbf\scriptfont\bffam=\sevenbf%
\scriptscriptfont\bffam=\fivebf\def\bf{\fam\bffam\tenbf}%
\normalbaselineskip=12pt%
\ifnum\linespacingstep>1\multiplelines\fi
\setbox\strutbox=\hbox{\vrule height8.5pt depth3.5pt width0pt}%
\let\sc=\eightrm\let\big=\tenbig\normalbaselines\rm}

\def\twelvepoint{\def\rm{\fam0\twelverm}
\textfont0=\twelverm\scriptfont0=\eightrm\scriptscriptfont0=\sixrm
\textfont1=\twelvei\scriptfont1=\eighti\scriptscriptfont1=\sixi
\textfont2=\magnifiedtensy\scriptfont2=\eightsy\scriptscriptfont2=\sixsy
\textfont3=\tenex\scriptfont3=\tenex\scriptscriptfont3=\tenex
\textfont\itfam=\twelveit\def\it{\fam\itfam\twelveit}
\textfont\slfam=\twelvesl\def\sl{\fam\slfam\twelvesl}
\textfont\ttfam=\twelvett\def\tt{\fam\ttfam\twelvett}
\textfont\bffam=\twelvebf\scriptfont\bffam=\eightbf
\scriptscriptfont\bffam=\sixbf\def\bf{\fam\bffam\twelvebf}
\tt 
\normalbaselineskip=14pt
\ifnum\linespacingstep>1\multiplelines\fi
\setbox\strutbox=\hbox{\vrule height10pt depth5pt width0pt}
\let\sc=\eightrm\let\big=\twelvebig\normalbaselines\rm}

\twelvepoint

\vsize=22.6 true cm \hsize=17 true cm 
\clubpenalty=5000
\widowpenalty=5000

\hsize=17.7cm
\vsize=23.9cm

\font\headfont=cmti12
\font\titlefont=cmbx12 at 17.28pt
\font\authorfont=cmr17
\font\affilfont=cmti10
\font\lonefont=cmbx12
\font\ltwofont=cmbx12

\def\textpar{\nobreak\medskip\noindent\ignorespaces}

\def\title#1{\noindent {\titlefont #1}\par \bigskip}
\def\author#1{\noindent {\authorfont #1} \par}

\long\def\abstract#1{\hbox to \hsize{\hfill \vbox{\hsize=13.5cm
\noindent{\lonefont \uppercase{Abstract}} \par \noindent #1 \par}}}

\def\sect#1{\par\goodbreak\bigskip
\noindent{\lonefont\nextsect~ \uppercase{#1}}\par\textpar}

\def\subsect#1{\par\goodbreak\bigskip
\noindent{\ltwofont\nextsub~ #1}\par\textpar}

\def\sectandsub#1#2{\par\goodbreak\bigskip
\noindent{\lonefont\nextsect~ \uppercase{#1}}\par\nobreak\vskip 1ex
\noindent{\ltwofont\nextsub~ #2}\par\textpar}

\newcount\sectcount
\newcount\subcount
\newcount\ssubcount
\def\nextsect{\global\advance\sectcount by 1
        \number\sectcount \global\subcount=0}
\def\nextsub{\global\advance\subcount by 1 \number\sectcount.\number\subcount
        \global\ssubcount=0}
\def\nextssub{\global\advance\ssubcount by 1
\number\sectcount.\number\subcount.\number\ssubcount}
\sectcount=0

%
%
\newcount\eqcount
\eqcount=0
\def\equno{\global\advance\eqcount by 1 \number\eqcount}
\def\equat#1{\count99=\eqcount \advance\count99 by #1 \number\count99}
\def\eqnam#1{\xdef#1{\the\eqcount}}

%
%
\newcount\figcount
\figcount=0
\def\nextfig{\global\advance\figcount by 1 Figure~\number\figcount}
\def\figno#1{\count99=\figcount \advance\count99 by #1 Figure~\number\count99}
\def\fignam#1{\xdef#1{\chaphead\the\figcount}}

%
%
\newcount\notecount
\notecount=0
\def\note#1{\global\advance\notecount by 1
   \footnote{$^{\the\notecount}$}{\tenpoint #1\par}}

\def\caption#1{\par\narrower\tenpoint\noindent{\bf Figure #1.}}

\def\refs{\parskip=0pt\par\goodbreak\blankline
\parindent=0pt\everypar{\hangindent 1cm}
\lonefont References\par
\nobreak\vskip 6pt\tenpoint

\def\AAp{{\it Astr. Astrophys.,} }
\def\AApS{{\it Astr. Astrophys. Suppl.,} }
\def\AJ{{\it Astr. J.,} }
\def\AnnRev{{\it Ann. Rev. Astr. Astrophys.,} }
\def\ApJ{{\it Astrophys. J.,} }
\def\ApJL{{\it Astrophys. J. Lett.,} }
\def\ApJS{{\it Astrophys. J. Suppl.,} }
\def\ApSS{{\it Astrophys. Sp. Sci.,} }
\def\BAN{{\it Bull. astr. Inst. Netherlands,} }
\def\JCP{{\it J. Comp. Phys.\/} }
\def\MNRAS{{\it Mon. Not. R. astr. Soc.,} }
\def\Kluwer{(Dordrecht: Kluwer)}
\def\Messenger{{\it ESO Messenger} }
\def\Nature{{\it Nature\/} }
\def\Obs{{\it Observatory,} }
\def\PASJ{{\it Publs astr. Soc. Japan,} }
\def\PASP{{\it Publs astr. Soc. Pacif.,} }
\def\PhD{{\it PhD thesis,} }
\def\PhilTrans{{\it Phil. Trans. R. Soc. Lond. A,} }
\def\PhysFl{{\it Phys. Fluids,} }
\def\PhysRep{{\it Phys. Reports} }
\def\Reidel{(Dordrecht: Reidel)}
\def\RMP{{\it Rev. Mod. Phys.} }
\def\RPP{{\it Rep. Prog. Phys.} }
\def\SovAst{{\it Soviet Astr.} }
\def\SovAstL{{\it Soviet Astr. Letters} }
\def\Vistas{{\it Vistas Astron.} }

}
\input psfig.sty

\overfullrule=0pt

\linespacingstep=1 \multiplelines \normalbaselines

\footline{\hfil}
\headline{\ifnum\pageno=1
    \hfil Revised version to appear in {\it Monthly Notices\/}
  \else
    {\headfont Instabilities of oblate spheroids} \hfil \tenrm\folio \hfil
  Sellwood \& Valluri
  \fi}
\raggedbottom

\newbox\topbox
\long\def\mtopbox#1{\setbox\topbox=\vbox{#1}
  \topinsert{\box\topbox}\endinsert}

\mtopbox{\vskip 0.5cm
\line{\titlefont Instabilities of a family of oblate stellar spheroids \hfil}}

\bigskip
\line{\authorfont J. A. Sellwood$^1$ and M. Valluri$^{1,2}$ \hfil}
\line{\affilfont $^1$Department of Physics and Astronomy, Rutgers
University, Piscataway, NJ 08855 \hfil}
\vskip-3pt
\line{\affilfont $^2$Department of Astronomy, Columbia
University, New York, NY 10027 \hfil}

\bigskip
\line{Rutgers Astrophysics Preprint No 193 \hfil}

\bigskip
\abstract{We have examined the stability of a sequence of oblate elliptical
galaxy models having the St\"ackel form suggested by Kuz'min \& Kutuzov.  We
have employed the 2-integral \DF s given by Dejonghe \& de Zeeuw for which
flattened non-rotating models are characterized by counter-streaming motion
and are radially cool; we introduce net rotation in some models by changing
the sign of the $z$-component of angular momentum for a fraction of the
particles.  We have found that all non-rotating and slowly rotating members
of this sequence rounder than E7 are stable, and that even maximally rotating
models rounder than E4 are stable.

In the absence of strong rotation, the most disruptive instability, and the
last to be stabilized by increasing thickness, is a lopsided ($m=1$) mode.
This instability appears to be driven by counter-rotation in radially cool
models.  Its vigour is lessened as rotation is increased, but it remains
strong even in models with net angular momentum 90\% of that of a maximally
rotating model before finally disappearing in maximally rotating models.
Strongly rotating models are more unstable to bar-forming modes which afflict
maximally rotating models with $c/a \ltsim 0.5$, but this mode is quickly
stabilized by moderate fractions of counter-rotating particles.  Bending
instabilities appear not to be very important; they are detectable in the
inner parts of the flatter models, but are less vigorous and more easily
stabilized than the lop-sided or bar modes in every case.  We briefly discuss
the possible relevance of the lop-sided instability to the existence of many
lop-sided disk galaxies.

}

\bigskip

\sect{Introduction}
The instabilities of equilibrium stellar systems deserve attention for at
least three reasons: systems known to be unstable can be excluded as possible
models of galaxies, knowledge of instabilities will help us to understand how
the galaxies can be altered during and after their formation into the shapes
we observe today, and finally, features such as bars, spirals, boxiness and
more speculatively lopsidedness, could result from instabilities.

In contrast to disk galaxies comparatively little work has yet been done on
the possible instabilities of realistic pressure supported stellar systems.
One of the principal reasons for this imbalance is that it is more difficult
to construct good equilibrium models of elliptical galaxies to study.  The
rotational support of disk galaxies, and the ready identification of the only
two possible isolating integrals in razor-thin disks, has allowed many models
to be constructed and their stability to be examined (see \eg\ Sellwood 1994
for a review).

\subsect{Historical survey}
Antonov (1960) used a variational principle to establish a criterion for the
Jeans stability of a spherical system.  Subsequent work on spherical systems,
nicely summarized by Binney \& Tremaine (1987, \S5.2), has yielded stability
criteria for all modes if the distribution function (\DF) is a function of
energy only (isotropic models), or to radial modes only for more general \DF
s.

Lynden-Bell (1967) formulated the normal mode problem for inhomogeneous
models, reworking Antonov's results for spherical systems and also discussing
the possible stellar dynamical equivalent of the two-stream instability of
plasmas.  Kalnajs (1971) introduced action-angle variables and developed a
matrix method (Kalnajs 1977) for searching for normal modes which has so far
been applied in rather few cases.

The simple form of the gravitational potential in uniform density spheroids
permits semi-analytic derivations of their normal modes (Fridman \&
Polyachenko 1984; Vandervoort 1991).   Some of these modes have analogues in
more realistic inhomogeneous models, but others are likely to be artifacts of
this sharp-edged mass distribution in which all orbits have an identical set
of frequencies.

May \& Binney (1986) and Goodman (1988) adopt a third approach of imagining a
neutral disturbance to the equilibrium and asking whether the system
over-responds; this yields a stability test, but not the form or vigor of any
instability, and is generally restricted to non-rotating models.

The final possible approach, which has yielded the most significant results,
is to test the stability of models using $N$-body simulations.  There are
many advantages to this approach: arbitrarily complicated models can be
studied without restriction on the geometrical shape or form of the \DF,
provided that an equilibrium model can be created.  Simulations reveal all
large-scale instabilities, without prior knowledge of their expected forms,
and yield the non-linear evolution which could either be violently disruptive
or cause a mild re-arrangement.  The physical behaviour can easily be
restricted, \eg\ by imposing symmetries, in order to separate simultaneously
growing modes or to obtain helpful insights into the instability mechanism.
The method comes with a number of disadvantages, of course: a simulation
yields only approximate frequencies and eigenfunctions of the first couple of
dominant modes.  Slowly growing modes may be masked by particle noise or
other inadequacies making it hard to determine the marginal stability point
in a sequence of models.  Finally, even large-scale instabilities can be
suppressed by quite modest amounts of gravity softening (\eg, Sellwood 1981;
1983).

Despite its generality, this approach has been most severely restricted by
the paucity of good equilibrium models to study.  With the exception of the
recent study by Dehnen (1996), the only non-spherical equilibria to have been
examined are shell orbit models (Bishop 1987, Hunter \etal\ 1990) or
numerically determined equilibria (Levison \& Richstone 1987).  In this paper
we use an analytic \DF\ for very simple St\"ackel models.  Recent progress in
constructing realistic equilibria (\eg\ Qian \etal\ 1995, Merritt \& Fridman
1996) should improve this situation dramatically in the near future.

\subsect{Known Instabilities}
Instabilities that have been found so far fall into two broad categories: the
well-known Jeans modes, in which the self-gravity of the disturbance is
destabilizing, and bending modes.  Bending modes, which resemble the
Kelvin-Helmholtz instability in fluids or the hose instability of plasmas
(Toomre 1966), are destabilized by anisotropic pressure and gravity actually
provides the restoring force.

The radial orbit instability was first discussed by Antonov (1972) who argued
that spherical stellar systems with a large fraction of predominantly radial
orbits would be unstable to bar formation.  Merritt (1987) gives an excellent
review of the early work and relates the instability to Lynden-Bell's (1979)
orbit alignment mechanism for bars in disks.  Since this mechanism still
relies on the self-gravity of the orbits, it is a Jeans-type instability.
Subsequently, Saha (1992) used Kalnajs's (1977) matrix method to study the
stability boundary for various non-singular models, but could find no general
stability criterion.  Palmer, Papaloizou \& Allen (1990) showed that
axisymmetric spheroidal models with a bias towards radial orbits were also
unstable.

de Zeeuw \etal\ (1983) found that non-rotating, radially cold ``shell orbit''
models were unstable to clumping into ring ($m = 0$) modes when the mass
distribution was flatter than E5.8 while Merritt \& Stiavelli (1990) gave the
critical boundary as E6.  These modes are clearly the generalization of the
well-known axisymmetric Jeans instability of stellar disks (Toomre 1964).

Lop-sided ($m = 1$) instabilities were also present in {\it all\/} the oblate
models studied by Merritt \& Stiavelli and in similar counter-streaming
models reported by Levison, Duncan \& Smith (1990, hereafter LDS).  Zang \&
Hohl (1978) had previously discovered $m=1$ instabilities in counter-rotating
disks.  Merritt \& Stiavelli suggested that flattened stellar systems need a
minimum amount of radial kinetic energy to be stable to this mode, a
conjecture supported by Sellwood \& Merritt's (1994, hereafter SM) results
for disks.  Robijn (1995) used the matrix method to study the unstable modes
of oblate Kuz'min-Kutuzov models having little radial pressure, and again
concluded that all oblate shell models are unstable to the lopsided mode.  He
found that adding net rotation did not affect the growth rate of the
instability very much, but that it was substantially reduced by increasing
the radial velocity dispersion.

Merritt \& Hernquist (1991) found that strongly prolate models, again having
only shell orbits with no net rotation, were unstable to bending modes.  Raha
\etal\ (1991) found a similar instability in a rapidly rotating triaxial bar,
and SM showed that hot disks generally buckle when too thin.  These modes are
similar to the bending modes of stellar sheets first studied by Toomre (1966)
-- see Merritt \& Sellwood (1994) for a more complete reference list.

LDS and SM also found that highly flattened systems with strong
counter-rotation can develop two oppositely rotating bars.  The latter
authors conjectured that the instability which leads to each bar was the
well-known bar-forming instability of disks, with the two oppositely rotating
systems coupled only weakly, if at all.  We question this interpretation here.

Some studies have included net rotation.  LDS showed that models as flat as
E6 were unstable to the formation of $m=2$ bar modes and associated spiral
modes.  They identified two primary causes for the development of a bar:
radial orbit instabilities which dominated in the radially hot non-rotating
models and the usual bar-forming (``streaming'') instability in rotating
models.  They found that both $m = 2$ and $m = 1$ modes were present in
models with net rotation and they claimed, somewhat surprisingly, that the
growth rate of the $m=1$ mode was independent of the degree of rotation.  We
do not confirm this last result.  Dehnen (1996) found bar instabilities in
oblate ``cuspy'' models with net rotation.

Allen, Palmer \& Papaloizou (1992) claimed that nearly spherical models that
were made to rotate by flipping orbits were unstable to forming a triaxial
shape, but it now appears that this result was an artifact.  Similar models
reported here seem quite stable.

\bigskip
In this paper, we present a reasonably complete study of the oblate family of
St\"ackel models described by Kuz'min \& Kutuzov (1962) with the two-integral
\DF s given by Dejonghe \& de Zeeuw (1988).   They are still unrealistically
simple in comparison with elliptical galaxies, \Ignore{which are expected to
be functions of three independent integrals,}but they are an improvement over
many of the studies reported above because they are neither spherical, nor
uniform density, nor are they shell orbit models.

\sectandsub{Technique}{The Kuz'min-Kutuzov Model}
Kuz'min \& Kutuzov (1962) describe a set of inhomogenous axisymmetric mass
models of St\"ackel form.  Dejonghe \& de Zeeuw (1988) derived several \DF s
for these models, but we have restricted our study to the stability of their
two-integral oblate models, which for brevity we refer to as the KKDZ models.

\mtopbox{
\centerline{\psfig{figure=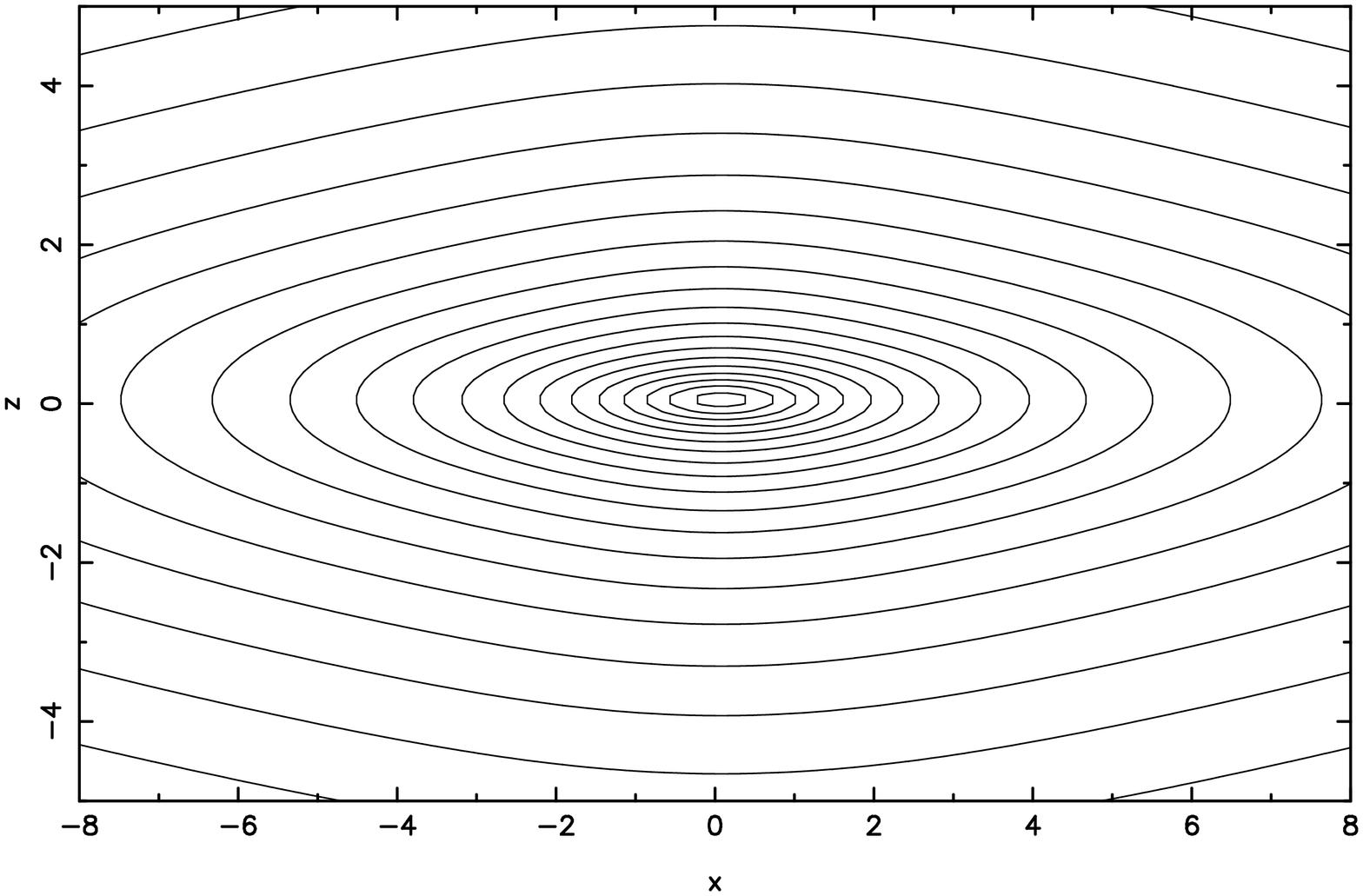,width=0.90\hsize,clip=,angle=270}}
\caption{1} Contours of projected density of the $c/a=0.2$ model, viewed from
the equatorial plane.}

In cylindrical polar coordinates, the models have the potential $$
\Phi(R, z) =  {GM \over (R^2 + z^2 + a^2 + c^2 + 2\sqrt{a^2 c^2 + c^2 R^2 +
a^2 z^2})^{1/2}}, \eqno(\equno)
$$ and density $$
\rho(R,z) = {Mc^2 \over 4\pi} {(a^2 + c^2) R^2 + 2 a^2 z^2 + 2 a^2 c^2 + a^4
+ 3 a^2 \sqrt{a^2 c^2 + c^2 R^2 + a^2 z^2} \over (a^2 c^2 + c^2 R^2 + a^2
z^2)^{3/2} (R^2 + z^2 + a^2 + 2 \sqrt{a^2 c^2 + c^2 R^2 + a^2 z^2})^{3/2}},
\eqno(\equno)
$$ where $M$ is the total mass.  Models with $c < a$ are oblate and models
with $c > a$ are prolate.  The family includes the spherical isochrone
(H\'enon 1959) when $a = c$, and the Kuz'min-Toomre disk (Kuz'min 1956,
Toomre 1963 model 1) when $c = 0$.

The surfaces of constant density are flattened spheroids and models become
more spherical with increasing radius (Dejonghe \& de Zeeuw 1988, Figure 5);
the density falls off as $r^{-4}$.  The fiducial axis ratio $c/a$ is
typically slightly smaller than the central axis ratio in the oblate case and
slightly greater than the central axis ratio in the prolate case.  The true
axis ratio approaches the limiting value, $\xi_{\infty}$, given by $$
\xi^4_{\infty} = {2c^3 \over a(a^2 + c^2)}. \eqno(\equno)
$$ As all our models are truncated at $R = 5(a+c)$, the outer axis ratio is
slightly less than the asymptotic value.

The {\it projected\/} density profile of the $c/a = 0.2$ model, viewed from
the symmetry plane to show maximum flattening, is shown in \nextfig.  We will
show in \S3 that this approximately E7 galaxy model is the flattest stable
member of the KKDZ family, except for maximally rotating models.

Dejonghe \& de Zeeuw applied inversion methods to derive a two-integral \DF\
for these spheroids.  For a given $c/a$, they find $$ \eqalign{
& F(E,J_z^2) = {1 \over 2^{3/2}\pi^3} {c^2 \over a} E^{5/2} \times \cr
& \qquad \left[ \sum_{\epsilon = -1,1} \int_0^1 {1-t^2 \over (1 -
2aEt\sqrt{1-t^2} + \epsilon t\sqrt{z})^5} \left[ (3+4x_\epsilon -
x_\epsilon^2) (1-x_\epsilon) (1-t^2) + 12t^2 \right] dt \right], \cr}
\eqno(\equno a)
$$ where, $$
x_\epsilon = {2aEt\sqrt{1-t^2} \over 1 + \epsilon t \sqrt{z}}, \qquad z =
2AEJ_z^2, \qquad A = {a^2-c^2 \over a^2}. \eqno(\equat0b)
$$ (Two typographical errors in the original published expression have been
corrected here.)  This \DF\ is nowhere negative for $c < a$ and is therefore
physical for all oblate models.

\mtopbox{
\centerline{\psfig{figure=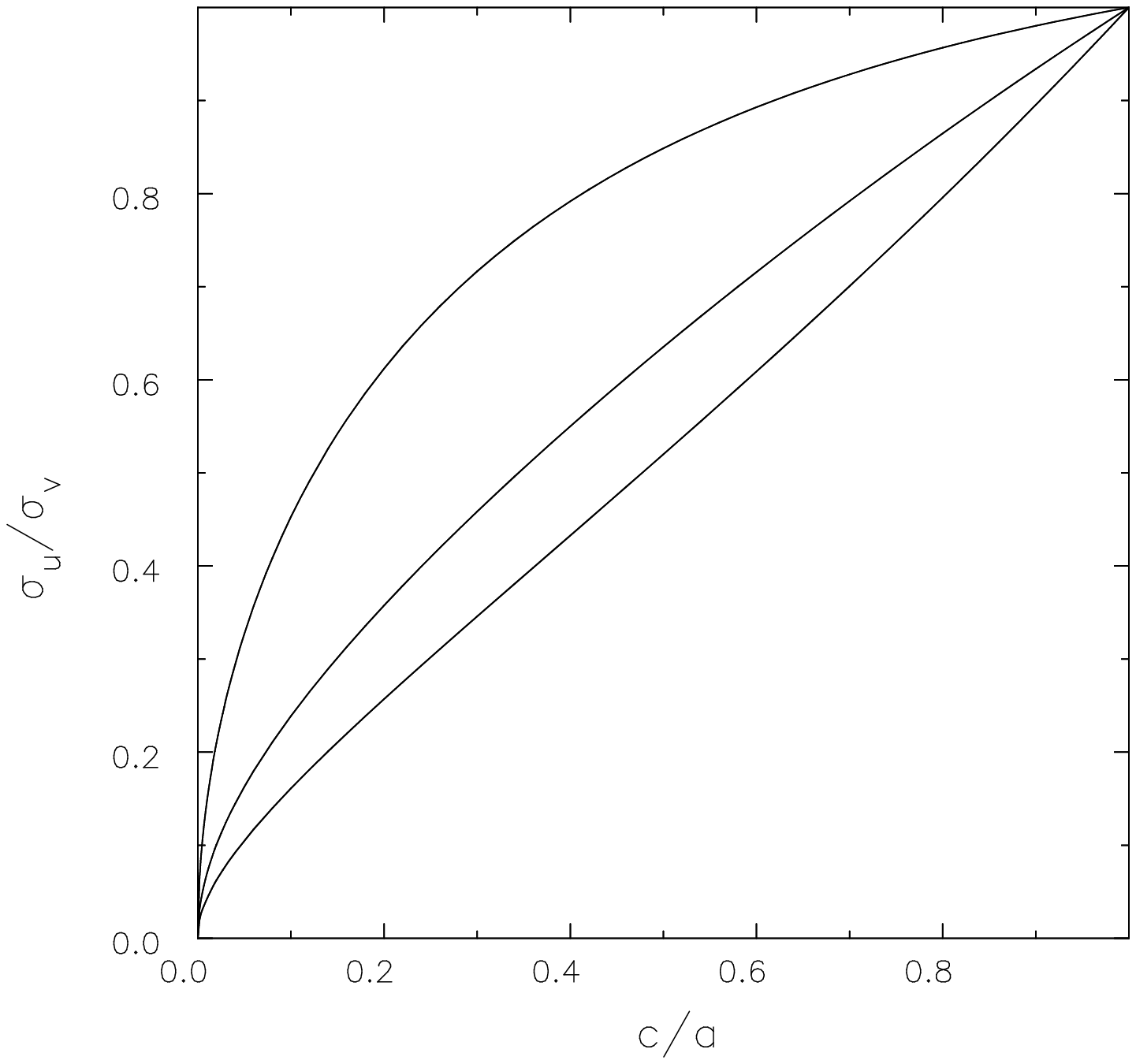,width=0.50\hsize,clip=}}
\caption{2} The axis ratio of the velocity spheroid in the equatorial plane
as a function of flattening.  From top to bottom, the three curves show
values at $R=0.5, 1.3$ \& 3.0 respectively.}

Since these are two-integral models, the radial, $\sigma_u$, and vertical,
$\sigma_w$, velocity dispersions are always equal, and flatter models have
less radial pressure.  Flat models with no net rotation are characterized by
equal and opposite streams of particles; \nextfig\ shows the variation of
$\sigma_u/\sigma_v$ with $c/a$, with $\sigma_v$ defined as the rms velocity
of this counter-streaming distribution at three different radii in the
equatorial plane: $(R,z) = (0.5,0)$, (1.3,0), where the streaming velocity
dispersion is close to maximum, and (3,0).  The anisotropy of the velocity
distribution increases with radius for flat models but becomes isotropic
everywhere for the sphere.

To make the models finite, we eliminate all particles having an energy
greater than the gravitational potential in the mid-plane at $R=5(a+c)$.  We
ignore the small imbalance this truncation introduces since the initial
virial ratios $T/|W|$ are within a few percent of the equilibrium value of
0.5.

We adopt units where $G = M = a + c = 1$.  The central potential $\Phi(0,0) =
1$ in these units.

\subsect{Quiet Start}
We generate particles by selecting pairs of $(E, J_z)$ values having the
density of this \DF\ using a technique similar to that described by Sellwood
\& Athanassoula (1986).  It ensures that the pairs of integrals of our
selected particles have the smoothest possible distribution for a discrete
sample drawn from the desired 2-integral \DF.  Having selected the integrals
deterministically, we choose the initial orbital phases randomly, which we
justify by appealing to past experience with disks.  Sellwood (1983) and Earn
\& Sellwood (1995) found that growth rates of non-axisymmetric instabilities
were not much affected by how carefully the angles were chosen.\note{In this
3-D St\"ackel model, however, we might have obtained more precise results had
we enforced a uniform density in the third integral.}  For each pair of
integrals, we position one or more particles (up to 5) by choosing $(R,z)$
pairs from the allowed region of the meridional plane (where the probability
density is uniform) and an azimuthal phase for each.  The values of $E$ and
$J_z$ almost determine the velocity components at the chosen position -- all
that remains is to direct the component in the meridional plane with the help
of one further random number.

In order to determine the unstable modes of the model, in addition to
requiring that the particles are the best possible sample from the \DF, we
also require a distribution that is both at rest and perfectly centered on
our coordinate origin and that the seed amplitude of all unstable modes is
much lower than is possible with a random distribution of particles.  To
achieve this, we use each coordinate set generated as described in the
previous paragraph to place two particles in mirror symmetric positions about
the $z=0$ plane with oppositely directed vertical velocity components and
then place six such pairs of particles equally in azimuth.  This procedure
automatically centers the model and eliminates all three linear momentum
components; we can also eliminate net angular momentum, if desired, by
reversing the azimuthal components of half these particles.  Imposing
six-fold rotational symmetry reduces the seed amplitude of all low-order
non-axisymmetric modes, which include all the most important
instabilities.\note{A study of axisymmetric oscillations, on the other hand,
would require a different strategy -- one designed to suppress noise in the
radial density profile.}

\subsect{Rotation}
The equilibrium of an axisymmetric stellar system is unaffected when an
arbitrary fraction of the stars have their angular momentum about the
symmetry axis reversed.  We are therefore able to introduce azimuthal
streaming motion by using Lynden-Bell's (1962) daemon to reverse the sign of
$J_z$ for a fixed fraction of particles.  Following LDS, we define the
parameter $$
\eta = {\sum J_{z_i} \over \sum |J_{z_i}|}. \eqno(\equno)
$$ This parameter varies from zero for models with no net rotation to unity
when all particles orbit in the same sense.  When $\eta =0.5$, the system has
half the maximum possible angular momentum: 75\% of the particles are
orbiting in the direct sense and 25\% are retrograde.

It should be noted that an orbit flipping rule which takes no account of the
magnitude of $J_z$ will introduce a discontinuity in the \DF\ across $J_z=0$.
  Since Kalnajs (1977) warned that such a discontinuity will aggravate the
bar mode, it is better to taper the flipping rule so that no signs are
changed when $|J_z|$ is small, up to the full desired fraction for
$|J_z|>0.2\sqrt{GMa}$ say.  We find that tapering the discontinuity in this
way reduces the net angular momentum of a maximally streaming model by a
percent or so, but decreases the growth rate of bar mode by more than a
factor of two.

\subsect{Numerical details}
We use a 3-D polar grid-based $N$-body code which is a straightforward
generalization of Sellwood's (1981) 2-D code, with the central hole
eliminated. The grid planes are equally spaced in $z$ and we solve for the
gravitational field of the mass distribution using FFTs in the vertical and
azimuthal directions and by direct convolution in the radial direction.  The
method closely resembles that described by Pfenniger \& Friedli (1993) except
that we solve separately for the three components of the gravitational force
field and for the potential, which eliminates the need for numerical
differences of the potential to find the force components.  We adopt the
standard (Plummer sphere) softening prescription to prevent strong forces at
short range; the fixed softening length is comparable to the vertical spacing
of the grid planes but is smaller than the horizontal mesh cell dimensions at
large distances from the symmetry axis.

We use linear interpolation between the grid points both for mass assignment
and to determine accelerations.  We advance the motion of each particle using
simple time-centered leap-frog in Cartesian coordinates.

The grid in most of our simulations has 80 nodes in azimuth, 65 in the radial
direction and 225 vertically.  We vary the $z$-spacing of the grid planes in
order to ensure the best possible spatial resolution while keeping the whole
model within the grid boundaries: typically $c/10 < \delta z \sim c/5$ so
that the spacing (and softening length) is small compared with any scale on
which the density varies.  We use $1.2 \times 10^5$ particles and a time step
of $0.05\sqrt{(a+c)^3/GM}$.  Test runs with ten times the number of particles
produced essentially identical results.   Results did vary, however, with
changes to the softening length and vertical grid spacing.  Results using
finer grids were not significantly different, but we found that grids coarser
than our standard size generally led to lower growth rates.

We evaluate and save coefficients of a low-order spherical Bessel function
expansion of the particle distribution at frequent intervals throughout the
run.  We then use the mode fitting procedure described by Sellwood \&
Athanassoula (1986) to estimate eigenfrequencies of the most unstable modes
during the period of exponential growth.

Some models possessed several different instabilities which grow at different
linear rates and saturate at different times.  Linear modes are decoupled
from each other at small amplitude, but all are affected as soon as one
begins to saturate.  The most rapidly growing mode is easily measured but it
may saturate before other instabilities have grown sufficiently to yield
reliable eigenfrequencies.  We can, however, study the modes individually by
re-running the same initial conditions with the disturbance forces restricted
to a single non-axisymmetric Fourier harmonic.  The Jeans or bending mode of
that single harmonic develops separately in each run, prolonging the linear
growth period for the milder instabilities and allowing us to measure the
growth rates of more slowly growing instabilities.  It is also possible to
impose reflection symmetry about the mid-plane to suppress the bending mode
when both Jeans and bending instabilities of the same azimuthal symmetry are
present.  The highly flattened models were unstable to axisymmetric Jeans
modes, which can be suppressed by setting the axisymmetric field components
to the analytic expressions.

\sect{Results}
In this section we summarize the results of our study of the KKDZ models.
Our chief objectives are to establish the stability boundaries in
non-rotating models, to determine the influence of rotation, and to search
for possible instabilities in models rounder than E6.  The eigenfrequencies,
in units of $\sqrt{GM/(a+c)^3}$, of the non-axisymmetric instabilities in all
the models we have run are summarized in Tables 1-3.

\subsect{Axisymmetric modes}
The two-integral \DF\ naturally implies that $\sigma_u = \sigma_w$; therefore
thinner models are also radially cooler.  Those with axis ratios $c/a <0.09$
(corresponding to an asymptotic axis ratio of 0.2) were violently unstable to
the axisymmetric Jeans instability.  We have not estimated growth rates for
these modes, since they start from high amplitude and saturate quickly.  Our
stability boundary, $c/a=0.1$ or E7.7, is flatter than the E5.8 given by de
Zeeuw \etal\ (1983) and E6 suggested by Merritt \& Stiavelli (1990) for shell
orbit models.  This difference is hardly surprising since radial pressure,
which contributes to stability (Toomre 1964), rises as our models become less
flat.  These instabilities are unaffected by rotation.

\subsect{Lopsided Modes $(m=1)$}
As in the case of previous studies of models with strong counter-rotation
(\eg, Merritt \& Stiavelli 1990; LDS; SM), we found a non-rotating lop-sided
instability in the flatter models.  Rounder models with $c/a \gtsim 0.2$
required special care because despite having cancelled net momentum at the
start, a lateral motion of the model causes spurious growth in $m=1$
coefficients determined from a fixed center.  We could eliminate an apparent
slow instability by re-centering the distribution of particles every 20 steps.

\mtopbox{
\centerline{\psfig{figure=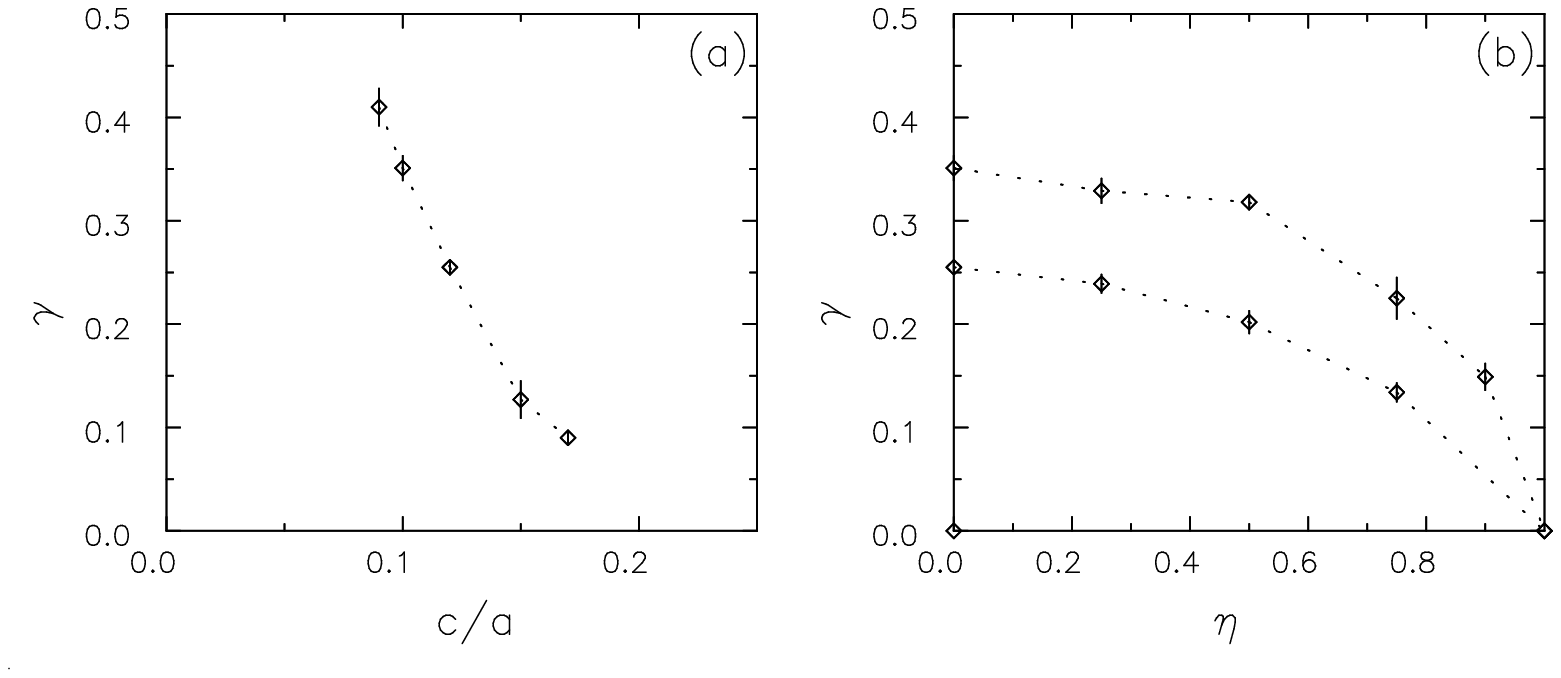,width=0.90\hsize,clip=}}
\caption{3} Growth rates of lop-sided modes in (a) non-rotating models and
(b) two sequences of rotating models with $c/a=0.1$ (upper) and $c/a=0.12$.
We have joined the measured values with dotted lines merely to indicate
sequences.}

\nextfig(a) shows that the growth rate, $\gamma$, of this mode decreases with
increasing $c/a$.  $N$-body methods are not well suited to pinpointing a
stability boundary with accuracy, but both the trend from the growth rates
and the fact that we could not detect an instability in models with $c/a \geq
0.2$ strongly suggests a stability boundary of the lop-sided mode lies at a
fiducial axis-ratio of $c/a \simeq 0.2$, which corresponds to an E7 galaxy
(Figure 1).

Merritt \& Stiavelli and Robijn (1995) had found that even slightly oblate
models are lop-sided unstable, but their models had little or no radial
pressure while the radial pressure in our two-integral models rises as they
become rounder.  Our result supports Robijn's finding that an increase in
radial pressure has a considerable stabilizing effect.

The lopsided instability persists in slowly rotating models, and disappears
only for maximally rotating cases.  The mode has a non-zero pattern speed and
resembles a one-armed spiral in models with intermediate values of $\eta$.
\figno0(b) shows the dependence of the growth rate on the degree of rotation
for two sequences of models with different axis ratios.  In both cases we
find that while $\gamma$ decreases as net rotation is increased, the
instability remains quite strong when $\eta$ is even as large as 0.9 and
disappears only when all the particles orbit in the same sense.  Note that,
LDS reported no dependence on the degree of rotation; the reason for this
difference is unclear.

\mtopbox{{\bf Table 1.} Growth rates of lop-sided instabilities
determined from the models

\blankline

$$\vbox{ \def\s{\kern\digitwidth} \halign{
\indent#\hfil &\quad#\hfil &\quad#\hfil &\quad#\hfil &\quad#\hfil
&\quad#\hfil &
\quad#\hfil \cr
$c/a$ & $\eta$ & growth rate & pattern speed \cr
\cr
0.09 & 0.00  &  $0.410\pm0.018$ \cr
0.10 & 0.00  &  $0.351\pm0.012$ \cr
0.12 & 0.00  &  $0.255\pm0.006$ \cr
0.15 & 0.00  &  $0.127\pm0.018$ \cr
0.17 & 0.00  &  $0.090\pm0.005$ \cr
0.20 & 0.00  &  undetectable    \cr
\cr
0.10 & 0.00 &  $0.351\pm0.012$                   \cr
0.10 & 0.25 &  $0.329\pm0.012$ & $0.115\pm0.003$ \cr
0.10 & 0.50 &  $0.318\pm0.005$ & $0.231\pm0.006$ \cr
0.10 & 0.75 &  $0.225\pm0.020$ & $0.297\pm0.011$ \cr
0.10 & 0.90 &  $0.149\pm0.013$ & $0.353\pm0.016$ \cr
0.10 & 1.00 &  undetectable                      \cr
\cr
0.12 & 0.00 &  $0.255\pm0.006$                   \cr
0.12 & 0.25 &  $0.239\pm0.009$ & $0.088\pm0.002$ \cr
0.12 & 0.50 &  $0.202\pm0.011$ & $0.162\pm0.009$ \cr
0.12 & 0.75 &  $0.134\pm0.009$ & $0.214\pm0.009$ \cr
0.12 & 1.00 &  undetectable                      \cr
\cr}}$$}

\nextfig\ shows the axis ratio of the velocity ellipsoid before and after the
instability saturates.  As in the case of the models studied by Merritt \&
Stiavelli this figure shows that the primary effect of the non-linear
evolution of the lopsided mode is an increase in the radial velocity
dispersion at the cost of the rms velocity of the counter-streaming motion.
Thus while the initial model was supported largely by azimuthal motions the
instability leads to one supported more by radial pressure.

We do not have a convincing mechanism for the lop-sided mode, though its
disappearance in maximally rotating models clearly indicates that
counter-rotation is responsible.  Comparison of our results with those of
Merritt \& Stiavelli and Robijn shows also that increasing radial pressure,
more than decreasing oblateness, stabilizes the mode.  Some analytic work has
been done for counter-rotation in disks: Palmer (1994, \S12.5) provides a
mathematical description in the WKBJ approximation which, however, does not
offer much physical insight.  Araki (1987) and Lovelace, Jore \& Haynes
(1996) argue that it could perhaps be the first known stellar dynamical
analogue of the two-stream instability of plasmas.

\subsect{Counter-rotating bars}
Unlike LDS and SM, our non-rotating models generally did not form pairs of
counter-rotating bars.  Some of our axisymmetrically unstable models
developed counter-rotating bar-like features after the ring instability
saturated, but when axisymmetric forces were held rigid and disturbance
forces confined to $m=2$, the model exhibited an $m=2$ bending mode (\S3.5)
only.  There were no hints of linearly growing, bi-symmetric, Jeans-type
disturbances -- the amplitudes of $m=2$ {\it density\/} variations merely
rose incoherently as the quiet start slowly disrupted.

We have traced the source of this initially puzzling discrepancy to a
difference in initial seed amplitude of the $m=2$ noise in the particle
distribution at the start.  LDS did not adopt any special precautions to
suppress particle noise, and the procedure adopted by SM imposed bi-symmetry,
thereby {\it raising\/} the initial amplitude of the $m=2$ Fourier component
above that expected from a random distribution of particles.  By contrast,
the particle distribution in our present simulations was made six-fold
symmetric (\S2.2), thus ensuring that $m=2$ components of the noise were very
weak at the start.

\mtopbox{
\centerline{\psfig{figure=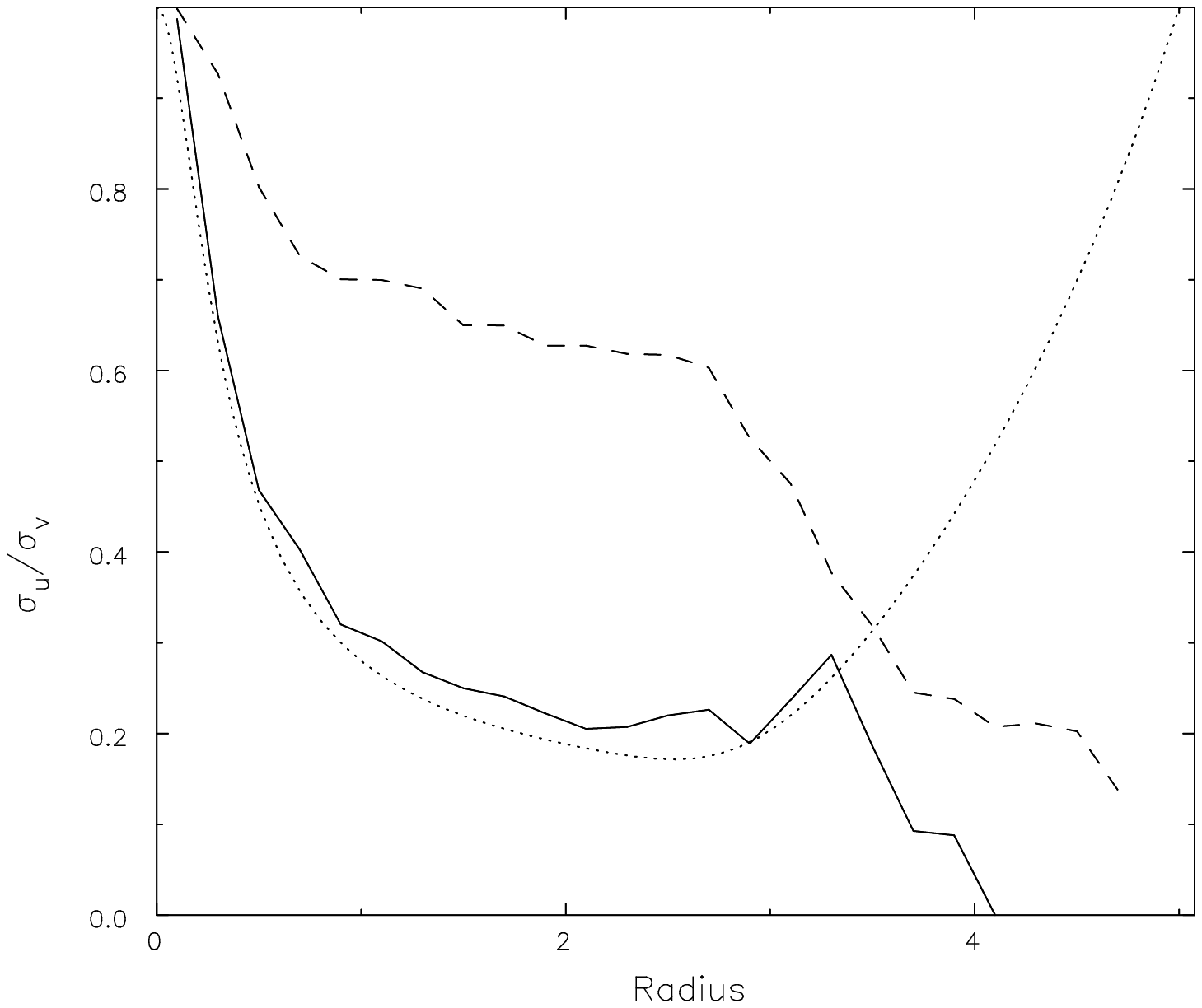,width=0.50\hsize,clip=}}
\caption{4} The axis ratio of the velocity ellipsoid before (full-drawn) and
after (dashed) the instability saturates.  The dotted curve shows the
expected value of this ratio in the mid-plane of this truncated model at the
start; the empirical values differ from the expected curve in the outer
region because the particle density is so low.}

We have verified in a number of ways that particle noise is the principal
source of the discrepancy.  The most direct was to test these models with the
initial bi-symmetric arrangement used by SM.  With this start, and with only
$m=2$ terms contributing to the disturbance forces, a pair of
large-amplitude, counter-rotating bars developed in a model with $c/a=0.8$
and weaker bars could also be detected in another model with $c/a=0.1$.

Our suspicions having been aroused, we used Sellwood's (1981) 2-D polar grid
code for a quick and inexpensive study of similar disks.  We examined the
stability of Kuz'min-Toomre disks with \DF s given by Kalnajs (1976) -- the
same family as used by Athanassoula \& Sellwood (1986) and as thickened disks
by SM.  We first studied a half-mass, cool KT disk with the counter-rotating
population replaced by rigid mass and found it to be quite stable when a
quiet start was employed, but a bar formed in a noisy start model.  The bar
appeared to form as a result of non-linear trapping caused by swing-amplified
spiral disturbances (\eg\ Toomre 1981) seeded by the random particle
distribution.  Even though noise seeds all Fourier components equally, and
the swing amplifier is effective for several low-order symmetries,
large-amplitude bars develop because trapping is much easier for $m=2$ due to
the slow variation of $\Omega-\kappa/2$ (see Binney \& Tremaine 1987
\S6.2.1.a).

When the rigid mass was replaced by a live population of counter-rotating
particles, both populations formed bars in this way, usually of similar but
not exactly equal strength and not always at the same time.  It is curious
that bars formed quite readily in a cool ($Q \sim 1$) disk, but not in a cold
disk; if all the orbits were precisely circular, and the axisymmetric part of
the force was held rigid, the spirals seeded by the noise were very tightly
wrapped and appeared unable to transport enough angular momentum to trap
particles into a bar -- at least until random motion had built up
significantly.  Note that SM also reported that no counter-rotating bars
developed in their cold disk.

Thus the conjectures by LDS and by SM that the counter-rotating bar pairs
they found were formed through linear instabilities seem unlikely to be
correct; the bars in fact result from non-linear orbit trapping in
finite-amplitude spiral disturbances.  Paradoxically, modest random motion
aggravates this finite-amplitude instability, although very hot systems are
again stable.  A finite-amplitude instability also makes it easier to
understand SM's finding that imposing reflection symmetry, thereby
suppressing the $m=2$ bending mode, considerably delayed the formation of the
counter-rotating bars.

\mtopbox{
\centerline{\psfig{figure=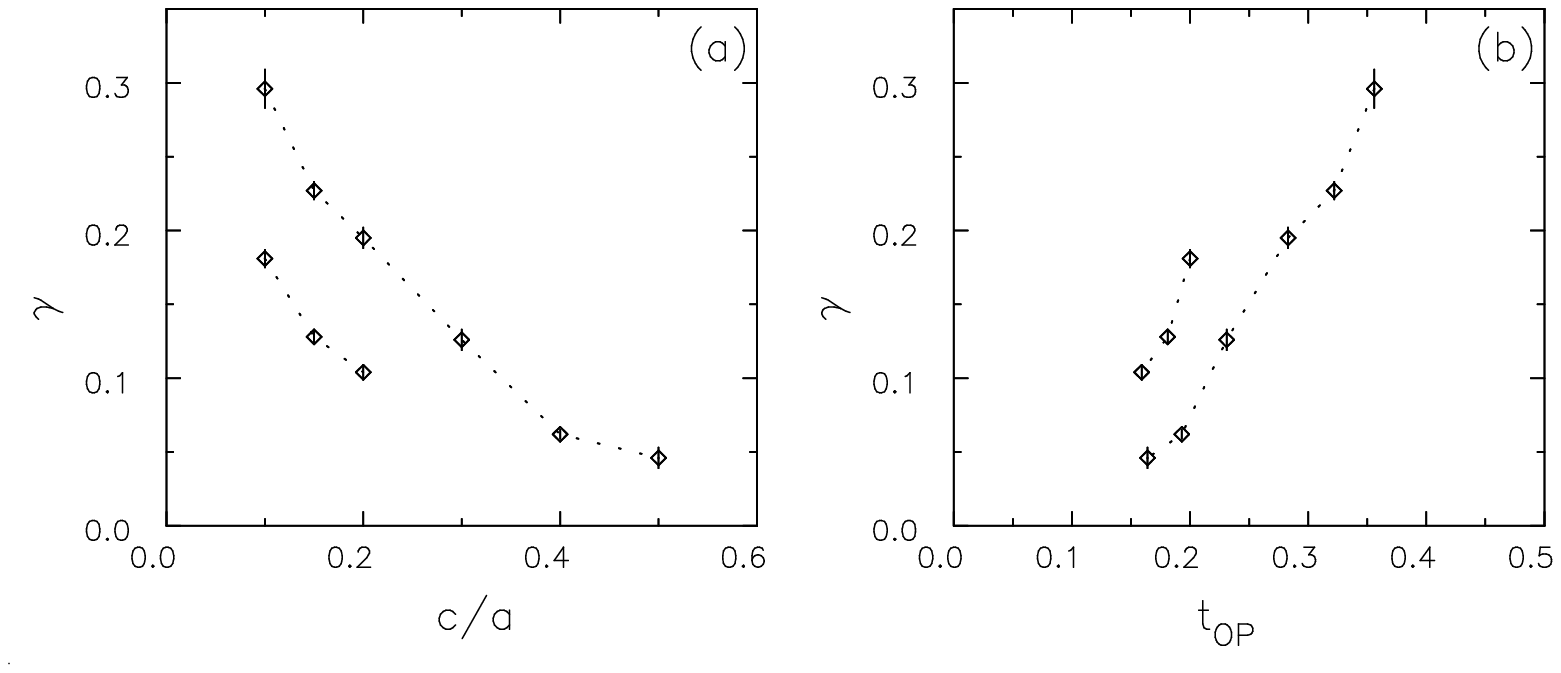,width=0.90\hsize,clip=}}
\caption{5} Growth rates of bar modes in rotating models with a smoothed \DF\
plotted as function of (a) axis ratio and (b) the Ostriker-Peebles $t$
parameter.  The longer sequence is for $\eta \sim 1$ while $\eta \sim 0.75$
for the shorter sequence.}

\mtopbox{{\bf Table 2.} Growth rates of bar instabilities determined from the
models

$$\vbox{ \def\s{\kern\digitwidth} \halign{
\indent#\hfil &\quad#\hfil &\quad#\hfil &\quad#\hfil & \quad#\hfil
&\quad#\hfil & \quad#\hfil \cr
 & &  & \multispan2 \hfil smoothed \DF \hfil & \multispan2 \hfil
discontinuous \DF \hfil \cr
$c/a$ & $\eta$ & $t_{\rm OP}$ & growth rate & pattern speed  & growth rate &
pattern speed \cr
\cr
0.10 &  0.992 & 0.356 & $0.296\pm0.013$ & $1.002\pm0.020$ & $0.695\pm0.005$ &
$1.471\pm0.029$ \cr
0.15 &  0.990 & 0.322 & $0.227\pm0.006$ & $0.830\pm0.008$ & $0.431\pm0.011$ &
$1.261\pm0.017$ \cr
0.2  &  0.989 & 0.283 & $0.195\pm0.007$ & $0.769\pm0.006$ & $0.364\pm0.023$ &
$1.111\pm0.016$ \cr
0.3  &  0.984 & 0.231 & $0.126\pm0.007$ & $0.671\pm0.004$ & $0.196\pm0.004$ &
$0.915\pm0.001$ \cr
0.4  &  0.979 & 0.193 & $0.062\pm0.003$ & $0.623\pm0.002$ & $0.139\pm0.001$ &
$0.838\pm0.00$3 \cr
0.5  &  0.974 & 0.164 & $0.046\pm0.007$ & $0.528\pm0.014$ & $0.092\pm0.006$ &
$0.756\pm0.005$ \cr
\cr
0.10 &  0.744 & 0.200 & $0.181\pm0.006$ & $0.880\pm0.004$ & $0.431\pm0.011$ &
$1.254\pm0.014$ \cr
0.15 &  0.743 & 0.181 & $0.128\pm0.004$ & $0.692\pm0.004$ & $0.199\pm0.009$ &
$1.052\pm0.017$ \cr
0.2  &  0.742 & 0.159 & $0.104\pm0.005$ & $0.635\pm0.005$ & $0.119\pm0.006$ &
$0.833\pm0.015$ \cr
\cr}}$$}

\subsect{Bar and Spiral Modes}
A bar instability, absent in quiet start non-rotating models, appears as
rotation is increased.  Table 2 gives two sets of frequencies for the bar
modes which are determined from models both with and without a discontinuity
in the \DF\ at $J_z=0$.  Smoothing the discontinuity, as described in \S2.3,
leads to the values of $\eta$ shown, while the values of $\eta$ for the
models with the discontinuity are $1.0$ and $0.75$ precisely; evidently the
discontinuity has a huge effect on the growth rates of bars (see Kalnajs
1977).

\nextfig\ shows our estimated growth rates for the $m=2$ mode from the
smoothed models as a function of axis-ratio for two values of the rotation
parameter $\eta$.  The growth rate drops rapidly both as the models become
rounder and as net rotation is reduced.  We could not detect any convincing
instabilities in models with rotation parameter $\eta \ltsim 0.5$ nor in any
a maximally rotating model rounder than $c/a=0.5$.  There was just a hint of
a coherent rotating wave in the maximally rotating model with $c/a=0.6$, but
we could not determine a credible growth rate and we therefore suspect this
model is stable, but only barely so.

The Ostriker-Peebles (1973) stability parameter, $t_{\rm OP}$, also given in
Table 2 for the smoothed models, appears to correlate well with our estimated
growth rates.  The trends in \figno0(b) suggest that models with $0.1<t_{\rm
OP} <0.15$ would have zero growth rates -- close to the suggested critical
value of $t_{\rm OP} \sim 0.14$.  It is no surprise that this stability
indicator works for our models, as it has been found to be serviceable for
all models with approximately harmonic cores.

The instability has the classic bi-symmetric spiral shape so often seen in
rotationally supported disks and is especially strong in flatter models.
While the swing-amplified feed-back loop described by Toomre (1981) seems to
account for many aspects of bar instabilities in thin disks, the picture of a
feed-back loop using thin-disk density waves would have to be stretched
unrealistically in the rounder members of our sequence, which are also
radially much hotter.  Sellwood's (1981; 1983) finding that surprisingly
little gravity softening could inhibit bar instabilities in a zero-thickness
disk suggests that thin-disk density waves are quickly weakened by increasing
thickness, but his result also contrasts with the persistence of bar-forming
instabilities to $c/a=0.5$ in the present models.  Clearly, a thick, hot
spheroid has different stability properties from a softened thin disk.   Thus
it seems possible the bar-forming instability in the thicker models may be
more closely related to the Riemann-type deformations of Maclaurin spheroids
(\eg, see Binney \& Tremaine 1987, \S5.3.5) while disk-like wave mechanics
may offer a better description of the mode in thinner systems.

The stability of our nearly spherical models contrasts with the findings of
Allen, Palmer \& Papaloizou (1992) who reported $m=2$ instabilities in nearly
spherical models with even small amounts of rotation.  To ensure that our
different result was not due to the different models used, we checked the
stability of their models, even using the same file of initial particle
coordinates (kindly supplied by A.\ J.\ Allen), but found their model to be
completely stable.  The discrepancy was eventually traced to an artifact in
their $N$-body code caused by sub-division of time steps without
recalculating the potential; they now concur that their rotating nearly round
models were stable.

\subsect{Bending Modes}
The only bending instability we found was a mild $m=2$ ``saddle'' mode in
highly flattened, slowly rotating, models.  This instability could be
reliably detected only in quite flattened models with little net rotation
where it was always dominated by the lop-sided mode.  It is clearly much less
vigorous that the bar mode in more rapidly rotating models, but we cannot
confirm its presence in this regime.

\nextfig\ shows the growth rates for various values $c/a$ for non- and slowly
rotating models, indicating that those flatter than $c/a = 0.17$ were
unstable.   In order to measure meaningful linear growth rates for this mode,
we have to suppress the lop-sided mode, by eliminating $m=1$ disturbance
forces, and axisymmetric disturbances in models with $c/a<0.1$.  At a fixed
axis ratio, the growth rate is reduced slightly by small net rotation, but
the bar instability prevents us from tracing the trend with increasing net
rotation any further.  Since this bending instability is driven by
counter-rotation, it presumably disappears for maximally rotating models.  We
found that the bends developed in the inner regions only where the models are
flatter (see Figure 1) and the radial extent of the mode decreased with
increasing thickness.

\mtopbox{
\centerline{\psfig{figure=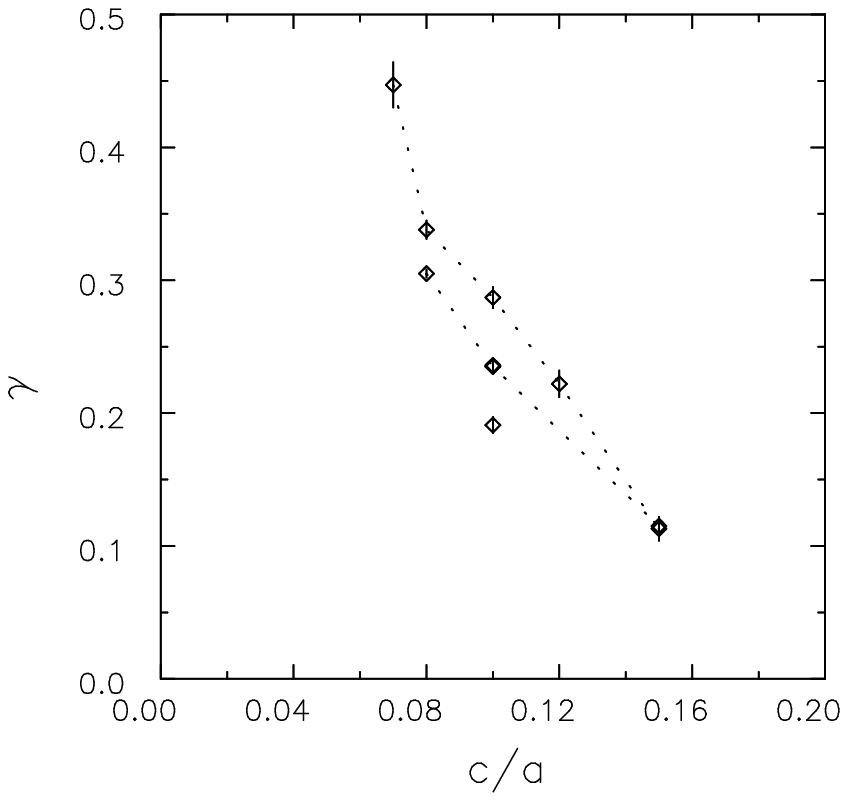,width=0.45\hsize,clip=}}
\caption{6} Growth rates of saddle modes plotted as function of axis ratio.
The two sequences are non-rotating ($\eta = 0$ upper) and slowly rotating
($\eta=0.25$).}

\mtopbox{{\bf Table 3.}  Growth rates of $m=2$ bending instabilities
determined from the models

$$\vbox{ \def\s{\kern\digitwidth} \halign{
\indent#\hfil &\quad#\hfil &\quad#\hfil &\quad#\hfil &\quad#\hfil
&\quad#\hfil &
\quad#\hfil \cr
$c/a$ & $\eta$ & growth rate & pattern speed \cr
\cr
0.07 & 0.00 &  $0.447\pm0.017$  \cr
0.08 & 0.00 &  $0.338\pm0.007$  \cr
0.10 & 0.00 &  $0.287\pm0.008$  \cr
0.12 & 0.00 &  $0.222\pm0.010$  \cr
0.15 & 0.00 &  $0.113\pm0.009$  \cr
0.17 & 0.00 &  undetectable  \cr
\cr
0.10 & 0.15 &  $0.236\pm0.001$ & $0.050\pm0.002$ \cr
0.08 & 0.25 &  $0.305\pm0.003$ & $0.106\pm0.003$ \cr
0.10 & 0.25 &  $0.235\pm0.005$ & $0.088\pm0.002$ \cr
0.15 & 0.25 &  $0.115\pm0.001$ & $0.055\pm0.004$ \cr
0.10 & 0.50 &  $0.191\pm0.006$ & $0.169\pm0.003$ \cr
\cr}}$$}

The mechanism for bending modes discussed by Merritt \& Sellwood (1994)
accounts pretty well for this instability in the present models.  Their
instability criterion requires that frequency of small-amplitude vertical
oscillations in the mid-plane exceed twice (for $m=2$) the circular
frequency.  The instability criterion was clearly satisfied in the radial
ranges where the instability was seen and is not satisfied anywhere when $c/a
\gtsim 0.2$.  However, their criterion marginally predicts instability when
$c/a=0.17$ whereas none was detected; the slight failure here is probably due
to the fact that the vertical oscillation period of nearly every particle
exceeds that in the exact mid-plane, and therefore too few particles could
co-operate to sustain an instability in this case.

\sect{Lop-sided Galaxies}
Many galaxies have some degree of asymmetry (\eg\ Richter \& Sancisi 1994;
Rix \& Zaritsky 1995): the most striking nearby cases include the LMC and
M101.  The origin of these asymmetries is still unclear -- not one of several
possible ideas has seemed sufficiently promising to have been worked out in
detail.  The asymmetries could be residuals of (late) galaxy formation, or
have been caused by recent interactions, or they could be long-lived $m=1$
density waves (Baldwin, Lynden-Bell \& Sancisi 1980; Earn 1993), or a weakly
damped mode (Weinberg 1994), or finally they could result from lop-sided
instabilities.

In principle, a dynamical instability would be the most attractive
explanation, since it would not require an external agent to excite it.
Unfortunately neither of the two known types of lop-sided instability seems
promising; as we have again shown, strong counter-rotation provokes a
lop-sided instability, but large fractions of counter-streaming stars are
believed to be rare (\eg, Kuijken, Fisher \& Merrifield 1996).  The other
lop-sided instability (Zang 1976; Sellwood 1985) requires a massive disk
surrounding a dense nucleus in which the rotation curve is flat or slightly
falling, but this does not seems promising either since many lop-sided
galaxies have rising rotation curves.

It occurs to us, however, that the absence of counter-rotation in the visible
matter does not preclude counter-rotation in the dark matter.  A strongly
flattened dark halo supported by counter-rotation could be lopsided unstable.
 Indeed, recent studies of the polar-ring galaxy NCG~4650A (Sackett \& Sparke
1990, Sackett \etal\ 1994) have concluded that the dark halo of this system
could be as flat as E6-E7 and Olling's (1996) study of flaring of the HI
layer in NGC~4244 also suggests a flattened halo.  On the face of it, we do
not think it likely that flattened dark halos would be supported by
counter-streaming, but the hypothesis could not be refuted by direct
observation.  In order to study the consequences for a low mass disk that
might happen to reside in a counter-rotating halo, we have introduced a disk
of test particles into one of our lop-sided unstable models.

\mtopbox{
\centerline{\psfig{figure=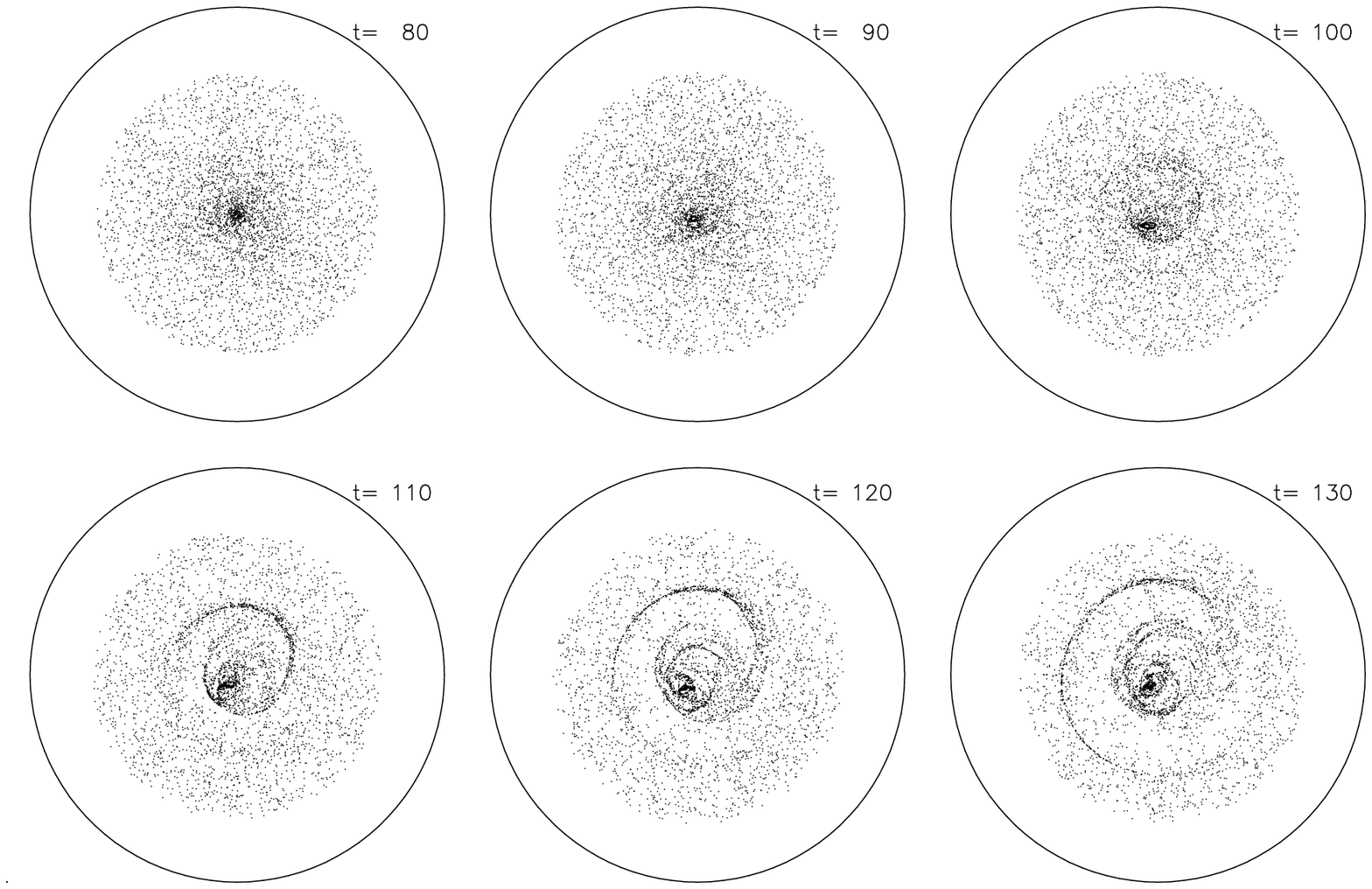,width=0.90\hsize,clip=,angle=270}}
\caption{7}  The evolution of a disk of test particles embedded in a ``dark
halo''.  A lop-sided instability in the halo begins to saturate at about time
100 and remains strong for some time.}

\nextfig\ shows part of the evolution of a disk of test particles lying in
the equatorial plane of the non-rotating $c/a = 0.15$ (approximately E7) KKDZ
model.  The test particles were given the appropriate orbital velocity (in
the anti-clockwise direction) to balance the central attraction from the KKDZ
model particles, which are not plotted.  The initial rotation curve of this
disk rises to a maximum at $R \simeq 1.0(a+c)$ and then declines slowly to
$\sim$ 85\% of the maximum rotation speed at the outer edge at $R = 4(a+c)$.
The response of this disk to the non-rotating lop-sided instability first
produces a lateral shift in the inner third of the disk and then a one-arm
leading spiral with an asymmetric density distribution in its interior.

The spiral arm, which is caused by a growing forcing amplitude, dissolves at
later times as the whole particle distribution becomes increasingly
disturbed.  We have chosen to determine the velocity field from the particles
at a moment when the lop-sidedness is strong but before the particle
distribution becomes too patchy.  \nextfig\ shows the 2-D velocity field of
the disk particles at time $t=120$ ``observed'' from two orthogonal
directions with the normal to the disk inclined at 30$^\circ$ to the line of
sight.  The extreme velocity contours close because the disk rotation curve
declines at large radii.  Even though the asymmetry in the mass distribution
appears quite small (Figure 7), the asymmetry in the velocity field is
striking.

\mtopbox{
\centerline{\psfig{figure=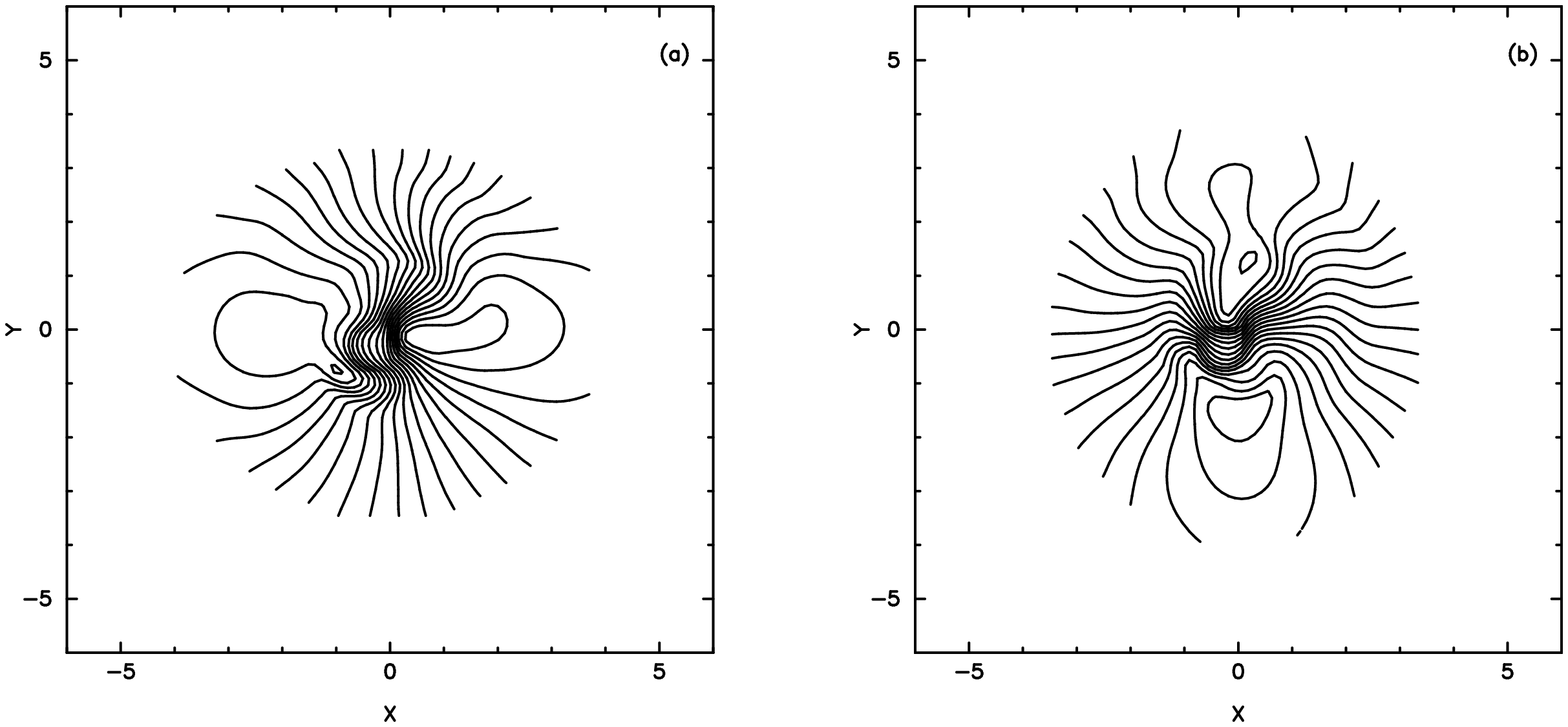,width=0.90\hsize,clip=,angle=270}}
\caption{8}  Two orthogonal projections of the velocity field of the disk of
test particles at $t=120$ with the disk inclined at $30^\circ$ to the line of
sight.}

While we do not think it likely that lop-sidedness in galaxies originates in
this way, the observable consequences of lop-sidedness in any dynamical model
are of some interest.  Our experiment also appears to illustrate that a mild
lop-sidedness in the mass distribution leads to more significant
lop-sidedness in the velocity field, and may perhaps help to account for the
very high incidence of asymmetries reported by Richter \& Sancisi from
kinematic data.

\sect{Conclusions}
The KKDZ family of models have proved to be quite remarkably stable, except
when highly flattened.  As the axis ratio is increased, the lop-sided mode is
the last to be stabilized in slowly rotating models, while the bar mode is
the last in maximally rotating models.  In the absence of extreme maximum
rotation, all models with $c/a \gtsim 0.2$ (asymptotic axis ratio 0.35)
appear to be completely stable.  We expect these stability properties are not
specific to this mass distribution, but are characteristic of any strongly
counter-streaming model (guaranteed by a 2-integral \DF\ for a flattened mass
distribution) having a uniform density core.

The addition of rotation has only a mild effect on the lop-sided mode unless
all counter-streaming particles are removed.  But as the net streaming motion
approaches its maximum, the bar instability sets in.  The sequence of
maximally rotating models is again stabilized by increasing thickness, but
not until $c/a \gtsim 0.5$ (asymptotic axis ratio 0.67), although less than
maximally rotating models are much more easily stabilized.  We have confirmed
Kalnajs's (1977) warning that the bar instability is strongly sensitive to a
discontinuity in the \DF\ across $J_z=0$; smoothing this discontinuity in a
reasonable manner reduces the net angular momentum by only a percent or two,
but more than halves the growth rate of the bar.

We do not confirm the bar-forming instabilities reported by Allen \etal\
(1992) in nearly round models with net rotation, which we have shown were a
numerical artifact of their code.  We have also shown that the
counter-rotating bars reported by LDS and by SM probably did not result from
linear instabilities, but were caused by non-linear trapping of particles in
large-amplitude disturbances seeded by particle noise.

The somewhat surprising stability of the slowly-rotating models seems to be
largely due to strong counter rotation.  The two principal instabilities
driven by counter-rotating streams of particles, the lop-sided and bending
modes, are quite quickly stabilized by radial and vertical pressure
respectively.  We expect more realistic models that are supported by a
slightly larger degree of radial pressure to be more unstable to disruptive
axisymmetric bending instabilities, as found by SM, while radial motions
strong enough to make the velocity ellipse radially biased would probably
destabilize the radial orbit mode.  We have chosen not to extend the present
study to include such models, which could be created using the 3-integral \DF
s discussed by Dejonghe \& de Zeeuw (1988), because it would be more
interesting to turn to models that bear a closer resemblance to elliptical
galaxies.

\bigskip
We would like to thank David Merritt for suggesting we undertake this study
and for the use of his adaptive Kernel algorithm for preparing Figure 8.
Thanks are also due to the referee, David Earn, for a careful read of the
paper.  This work was supported by NSF grant AST 93/18617 and NASA Theory
grant NAG 5-2803.

\sectcount=-1
\bigskip
{\refs
\def\bold{\bf}
\def\ital{\it}

Allen A.\ J., Palmer P., Papaloizou J., 1992, \MNRAS {\bold 256}, 695

Antonov V.\ A., 1960, {\ital Astr.\ Zh.}, {\bold 37}, 918 (English
translation: \SovAst {\bold 4}, 859)

Antonov V.\ A., 1972, in Omarov, T.\ B., ed, {\ital The Dynamics of Galaxies
and Star Clusters} (Alma Ata: Nauka of Kazakh SSR).  English translation in
de Zeeuw P.\ T., ed, IAU Symposium {\bold 127}, {\ital Structure and Dynamics
of Elliptical Galaxies} (1987), Reidel, Dordrecht, p.~549

Araki S., 1987, \AJ, {\bold 94}, 99

Athanassoula E., Sellwood J.\ A., 1986, \MNRAS {\bold 221}, 213

Baldwin J.\ E., Lynden-Bell D., Sancisi R., 1980, \MNRAS {\bold 193}, 313

Binney J., Tremaine S., 1987, {\ital Galactic Dynamics}, Princeton University
Press, Princeton

Bishop J.\ L., 1987, \ApJ {\bold 322}, 618

Dehnen W., 1996, preprint

Dejonghe J., de Zeeuw P.\ T., 1988 \ApJ {\bold 333}, 90

de Zeeuw P.\ T., Franx, M., Meys, J., Brink, K., Habing, H., 1983, in
Athanassoula, E., ed, IAU Symposium {\bold 100}, {\ital Internal Kinematics
and Dynamics of Galaxies}, Reidel, Dordrecht, p.~285

Earn D.\ J.\ D., 1993, \PhD Cambridge University

Earn D.\ J.\ D., Sellwood J.\ A., 1995, \ApJ {\bold 451}, 533

Fridman A.\ M., Polyachenko V.\ L., 1984, {\ital Physics of Gravitating
Systems}, Springer-Verlag, New York

Goodman J., 1988, \ApJ {\bold 329}, 612

H\'enon M., 1959, {\ital Ann.\ d'Astrophys}, {\bold 22}, 126

Hunter, C., de Zeeuw, P.\ T., Park, Ch., Schwarzschild, M., 1990, \ApJ {\bold
363}, 367

Kalnajs A.\ J., 1971, \ApJ {\bold 166}, 275

Kalnajs A.\ J., 1976, \ApJ {\bold 205}, 751

Kalnajs A.\ J., 1977, \ApJ {\bold 212}, 637

Kuijken K., Fisher D., Merrifield M.\ R., 1996 (preprint -- astro-ph/9606099)

Kuz'min G.\ G., 1956, {\ital Astr. Zh.}, {\bold 33}, 27

Kuz'min G.\ G., Kutuzov S.\ A., 1962, {\ital Bull Abastumani Ap Obs}, {\bold
14}, 52

Levison H.\ F., Richstone D.\ O., 1987, \ApJ {\bold 314}, 476

Levison H.\ F., Duncan M.\ J., Smith B.\ F., 1990, \ApJ {\bold 363}, 66 (LDS)

Lovelace R.\ E.\ V., Jore K.\ P., Haynes M.\ P., 1996, (preprint --
astro-ph/9605076)

Lynden-Bell D., 1962, \MNRAS {\bold 124}, 1

Lynden-Bell D., 1967, in Ehlers J., ed, {\ital Relativity Theory and
Astrophysics 2. Galactic Structure}, American Mathematical Society,
Providence RI, p.~131

Lynden-Bell D., 1979, \MNRAS {\bold 187}, 101

May A., Binney J., 1986, \MNRAS {\bold 221}, 13p

Merritt D., 1987, in de Zeeuw T., ed, IAU Symposium {\bold 127}, {\ital
Structure and Dynamics of Elliptical Galaxies}, Reidel, Dordrecht, p.~315

Merritt D., Fridman T., 1996, \ApJ {\bold 460}, 136

Merritt D., Hernquist L., 1991, \ApJ {\bold 376}, 439

Merritt D., Sellwood J.\ A., 1994, \ApJ {\bold 425}, 55

Merritt D., Stiavelli M., 1990, \ApJ {\bold 358}, 399

Olling, R.P., 1996, \AJ {\bold 112}, 481

Ostriker J.\ P., Peebles P.\ J.\ E., 1973, \ApJ 186, 467

Palmer P.\ L., 1994, {\ital Stability of Collisionless Stellar Systems},
Kluwer, Dordrecht

Palmer P.\ L., Papaloizou J., Allen A.\ J., 1990, \MNRAS {\bold 243}, 282

Pfenniger D., Friedli D., 1993, \AAp {\bold 270}, 561

Qian E.\ E., de Zeeuw P.\ T., van der Marel R.\ P., Hunter C., 1996, \MNRAS
{\bold 274}, 602

Raha N., Sellwood J.\ A., James R.\ A., Kahn F.\ D., 1991, \Nature {\bold
352}, 411

Richter O-G., Sancisi R., 1994, \AAp, {\bold 290}, L9

Rix H-W., Zaritsky D., 1995, \ApJ {\bold 447}, 82

Robijn F.\ H.\ A., 1995, \PhD Leiden University

Sackett P.\ D., Sparke L.\ S, 1990, {\bold 361}, 408

Sackett P.\ D., Morrison H.\ L., Harding P., Boroson T.\ A., 1994, \Nature,
{\bold 370}, 441

Saha P., 1992, \MNRAS {\bold 254}, 132

Sellwood J.\ A., 1981, \AAp {\bold 99}, 362

Sellwood J.\ A., 1983, \JCP {\bold 50}, 337

Sellwood J.\ A., 1985, \MNRAS {\bold 217}, 127

Sellwood J.\ A., 1994, in Franco J., et al., eds, {\ital Numerical
Simulations in Astrophysics}, Cambridge University Press, Cambridge p.~90

Sellwood J.\ A., Athanassoula E., 1986, \MNRAS {\bold 221}, 195

Sellwood J.\ A., Merritt D., 1994, \ApJ {\bold 425}, 530 (SM)

Toomre A., 1963, \ApJ {\bold 138}, 385

Toomre A., 1964, \ApJ {\bold 139}, 1217

Toomre A., 1966, in {\ital Geophysical Fluid Dynamics}, notes on the 1966
Summer Study Program at the Woods Hole Oceanographic Institution, ref.\ no.\
66-46, p.~111

Toomre A., 1981, in Fall S.\ M., Lynden-Bell D., {\ital Structure and
Evolution of Normal Galaxies}, Cambridge University Press, Cambridge, p.~111

Vandervoort P.\ O., 1991, \ApJ {\bold 377}, 49

Weinberg, M.\ D., 1994, \ApJ {\bold 420}, 597

Zang T.\ A., 1976 \PhD Massachusetts Institute of Technology

Zang T.\ A., Hohl F., 1978, \ApJ {\bold 226}, 521

}

\end